\newcounter{myctr}
\def\myitem{\refstepcounter{myctr}\bibfont\noindent\ifnum\themyctr>9\else\phantom{0}
\fi\hangindent17pt\themyctr.\enskip}

\documentclass{ws-ijqi}

\usepackage{amsmath}

\usepackage{amsfonts,amssymb}

\begin{document}

\markboth{Gennady P. Berman, Alexander I. Nesterov and Alan R. Bishop}
{Non-Hermitian Description of a Superconducting Phase Qubit Measurement}

\catchline{}{}{}{}{}

\title{NON-HERMITIAN DESCRIPTION OF A SUPERCONDUCTING PHASE QUBIT MEASUREMENT}

\author{ALEXANDER I. NESTEROV}

\address{Departamento de F{\'\i}sica, CUCEI, Universidad de Guadalajara,
Av. Revoluci\'on 1500\\
 Guadalajara,  Jalisco,  44420, M\'exico\\
nesterov@cencar.udg.mx}

\author{GENNADY P. BERMAN}

\address{Theoretical Division, Los Alamos National Laboratory\\
Los Alamos, New Mexico 87544, USA\\
gpb@lanl.gov}

\author{ALAN R. BISHOP}

 \address{The Associate Directorate for Theory, Simulation {$\&$} Computation (ADTSC), Los Alamos National Laboratory,
Los Alamos, NM 87544, USA \\
arb@lanl.gov}

\maketitle

\begin{history}
\received{Day Month Year}
\revised{Day Month Year}
\end{history}

\begin{abstract}
We present an approach based on a non-Hermitian Hamiltonian  to describe the process of measurement by tunneling of a phase qubit state. We derive simple analytical expressions which describe the dynamics of measurement, and compare our results with those experimentally available.
\end{abstract}



\bibliographystyle{unsrt}
\bibliographystyle{natbib}

\section{Introduction}

The superconducting phase qubit has significant potential applications in quantum information processing due to its convenient functional operations, its capability to reduce the effects of noise and decoherence,  and its future utilization in scalable quantum computers \cite{MNAU,MJM}. The improvement of the fidelity of the qubit state measurement still remains one of the main issues.

In this paper, we propose an approach based on non-Hermitian Hamiltonians to describe the dynamics of measurement of the superconducting phase qubit state \cite{MNAU,MJM,SAM,YHC,SMC,KNAB,NB1}. We obtain simple analytical expressions for the probability of  tunneling to the continuum. We compare our theoretical results with experiments on superconducting phase qubits \cite{SAM,YHC} and demonstrate that our theoretical predictions are in a good agreement with the available experimental data.

\section{The non-Hermitian approach to superconducting phase qubit measurement}

Non-Hermitian Hamiltonians naturally appear when the energy spectrum of a quantum system can be formally represented by both quasi-discrete (intrinsic) and continuous parts, and one performs a projection of the total wave function onto the  quasi-discrete part of the spectrum \cite{FU,RI,RI3,VZ,KND,KE}. In this case, the corresponding intrinsic energy levels acquire finite widths which are associated with transitions from the intrinsic states to the continuum. Then, the dynamics of the intrinsic states can be described by the Schr\"odinger equation with an effective non-Hermitian Hamiltonian \cite{KND,KE}.

We consider the current-biased qubit-continuum detector system based on a Josephson junction, which can be described by the Hamiltonian ($\hbar =1$) \cite{NB1}:
\begin{align}\label{eq1}
  {\mathcal H}_t=&\sum_{n}\omega_{n}|n\rangle \langle n|+ \sum_{m\neq n}\beta_{mn}(t)|m\rangle \langle n| +\int  E|E\rangle \langle E|dE
   +\sum _{n} \int  \alpha _{n} (E)|E\rangle \langle n|dE \nonumber \\
   &+\sum _{n} \int  \alpha^\ast _{n} (E)|n\rangle \langle E|dE  \; \;(m,n = 0,1),
\end{align}
where $\omega_n$ is the energy of level $|n\rangle$. The set of states $\{|n\rangle , |E\rangle \} $ forms a complete Hilbert space  basis, so that an arbitrary state vector $|\psi \rangle$ can be expressed as
 \begin{align}\label{eq2}
  |\psi \rangle = \sum_{n=0,1}C_{n}|n\rangle +  \int  C_E|E\rangle dE.
 \end{align}
We use the following notation for qubit states: $|0\rangle= |\downarrow \rangle$ and $|1\rangle= |\uparrow \rangle$. The non-vanishing matrix elements, $\beta_{mn}(t)$, are given by $\beta_{12}=\beta_{21}=\Omega _{0} \cos (\omega t+\varphi)$, where $\Omega _{0}$ is the on-resonance Rabi frequency and $\omega$ is the frequency of the biased ac current, $ I_{ac}(t)$. In general, the tunneling amplitudes, $\alpha _{n} (E)$, are complex and depend on the total energy.

Applying the Feshbach projection method \cite{NB1,RI,RI3,VZ}, we obtain an effective non-Hermitian Hamiltonian, ${\tilde {\mathcal H}}_{eff}=  {\mathcal H}- i \mathcal W$, which describes the dynamics of the qubit, where
\begin{align}\label{H0}
    {{\mathcal H}} = \left(\begin{array}{cc} {\omega_1} & {\Omega _{0} \cos (\omega t+\varphi )} \\ {\Omega _{0} \cos (\omega t+\varphi )} &\omega_0   \end{array}\right),
\end{align}
and
\begin{align}\label{eqh1}
  \mathcal W =\frac{1}{2}\left(
      \begin{array}{cc}
        \Gamma_1 & \Gamma_{01}\\
        \Gamma_{10} & \Gamma_0 \\
      \end{array}
    \right).
\end{align}
Here we consider the case in which the $\alpha _{n} $'s are real constants. Then, in terms of tunneling amplitudes, the rates of tunneling into the continuum are: $\Gamma _{0} =2\pi \alpha _{0}^{2} $, $\Gamma _{1} =2\pi \alpha _{1}^{2} $ and $ \Gamma_{10}=\Gamma _{01} = \sqrt{\Gamma _{0}\Gamma _{1}}$.

\begin{figure}[tbh]
\begin{center}
\scalebox{0.525}{\includegraphics{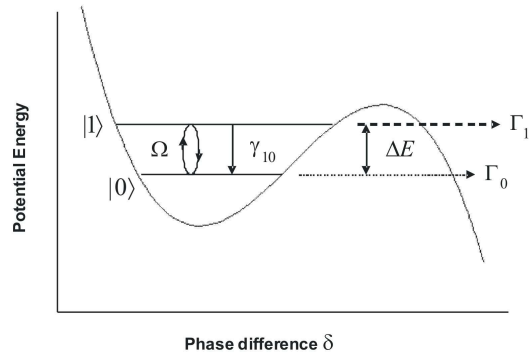}}
\end{center}
\caption{A superconducting phase qubit radiated by a microwave field; $\delta$ is the gauge-invariant phase difference of superconducting order parameter across the junction; $\gamma _{10} $ denotes the rate of energy relaxation from $|1\rangle $ to $|0\rangle $ and $\Gamma _{1} $($\Gamma _{0} $) denotes the tunneling rate from level $|1\rangle $($|0\rangle $).}
 \label{E}
\end{figure}

The state of the qubit is controlled  by a current $I(t)=I_{dc} -I_{\mu w} \cos (\omega t+\varphi ) $ with a dc bias current, $I_{dc}$, and a time-varying bias current, $I_{ac}=I_{\mu w}\cos (\omega t+\varphi )$, with frequency, $\omega $ \cite{SMC,MNAL}. The bias current, $I_{dc}$, can be chosen so that the tunneling is mostly from the upper level $|1\rangle $\cite{YHC}. (See Fig. \ref{E}.) In what follows, we study the dynamics of quantum measurement by tunneling of a superconducting phase qubit in the limit $\Gamma _{1} \gg \Gamma_0,\gamma _{10},\gamma _{\varphi}$, where $\gamma_\varphi$ is the dephasing rate. (We will discuss interaction with environment elsewhere \footnote{In a more complete model \cite{NB1}, we consider the dynamics of the system within the generalized master equation in the Lindblad form, $  \dot{ \rho} = -i[{\mathcal H},\rho] + L\rho -\{\mathcal W,\rho\}$, where $\{\mathcal W,\rho\}= \mathcal W\rho +\rho\mathcal W$ and $L\rho =  \frac{1}{2}\sum_{k=1}^{2}\{2  L_k \rho L^\dagger_k -L_k^\dagger  L_k\rho - \rho L_k^\dagger  L_k\}$, and transition operators, $L_k$, describe the coupling to the environment \cite{LG,CH}. }.)

\subsection{Rotating wave approximation}

In the rotating wave approximation, by removing fast oscillations, we find that ${\tilde {\mathcal H}}_{eff} \rightarrow {\tilde {\mathcal H}}$, where
\begin{align}\label{H3}
\tilde{ {\mathcal H}}=\frac{1}{2} \left(\begin{array}{cc} {\lambda_0+\Delta -i\Gamma_1}& {\Omega _{0} {\kern 1pt} e^{-i\varphi } } \\ {\Omega _{0} {\kern 1pt} e^{i\varphi }  } & {\lambda_0-\Delta -i\Gamma_0} \end{array}\right).
\end{align}
Here $\lambda_0 =\omega_{0} +\omega_{1}$; $\Delta = \omega_{10}  - \omega$ is the detuning (with $\omega $ being the microwave frequency; and we denote $\omega_{10}= \omega_{1} - \omega_{0}$). We neglect here the off-diagonal term $\Gamma_{01}$, which describe the interaction between the two tunneling channels. We let the wave function $|u(t)\rangle =C_1(t)|1\rangle +C_0(t)|0\rangle$ describe the qubit state. We assume below that $\langle u(0)|u(0)\rangle =1$. Then the solution of the Schr\"odinger equation with the non-Hermtian Hamiltonian (\ref{H3}) can be written as
\begin{align}\label{eq11a}
&C_{1}(t)  =e^{-i\tilde\lambda_0 t/2} \bigg(\Big(\cos\frac{\Omega t}{2} - i\cos\theta\sin\frac{\Omega t}{2}\Big) C_1(0)-  i e^{-i\varphi}\sin\theta\sin\frac{\Omega t}{2} C_0(0)\bigg),\\
&C_{0}(t)  = e^{-i\tilde\lambda_0 t/2} \bigg (\Big(\cos\frac{\Omega t}{2} +i\cos\theta\sin\frac{\Omega t}{2}\Big) C_0(0)-  i e^{i\varphi}\sin\theta\sin\frac{\Omega t}{2} C_1(0)\bigg ),
\label{eq11b}\\
&\rho_{11}(t)  = {e^{-\Gamma t}} \bigg |\Big(\cos\frac{\Omega t}{2} - i\cos\theta\sin\frac{\Omega t}{2}\Big) C_1(0)-  i e^{-i\varphi}\sin\theta\sin\frac{\Omega t}{2} C_0(0)\bigg |^2, \label{P3a}\\
&\rho_{00}(t)  = {e^{-\Gamma t}} \bigg |\Big(\cos\frac{\Omega t}{2} +i\cos\theta\sin\frac{\Omega t}{2}\Big) C_0(0)-  i e^{i\varphi}\sin\theta\sin\frac{\Omega t}{2} C_1(0)\bigg |^2,
\label{P3b}
\end{align}
where $\tilde\lambda_0= \omega_0 +\omega_1 -i \Gamma$, $\Gamma= (\Gamma_0 + \Gamma_1)/2$,  $\cos\theta = (\Delta - i(\Gamma -\Gamma_0))/\Omega$ and $\sin\theta = \Omega_0/\Omega$, the complex Rabi frequency is $\Omega= \sqrt{\Omega_0^2 +(\Delta - i(\Gamma -\Gamma_0))^2}$; the upper level population is $\rho_{11}(t)=|C_1(t)|^2$ and the lower level population is $\rho_{00}=|C_0(t)|^2$.

In what follows we compare our theoretical results with experiments on quantum state tomography of a superconducting phase qubit \cite{SAM}, and with the data obtained in the experiments observing coherent temporal oscillations between the quantum states of a Josephson tunnel junction \cite{YHC}. In both experiments the tunneling rate from level $|1\rangle$ is more than $10^3$ times that from level $|0\rangle$. So the total tunneling rate to the continuum is determined by the population of the upper level. The bias current was chosen so that the tunneling from level $|0\rangle$ is ``frozen out'' and escapes mostly from the upper level \cite{SAM,YHC}. Therefore, the time-dependent tunneling probability is defined by the upper level population $\rho_{11}(t)$.

\begin{figure}
\begin{center}
\scalebox{0.275}{\includegraphics{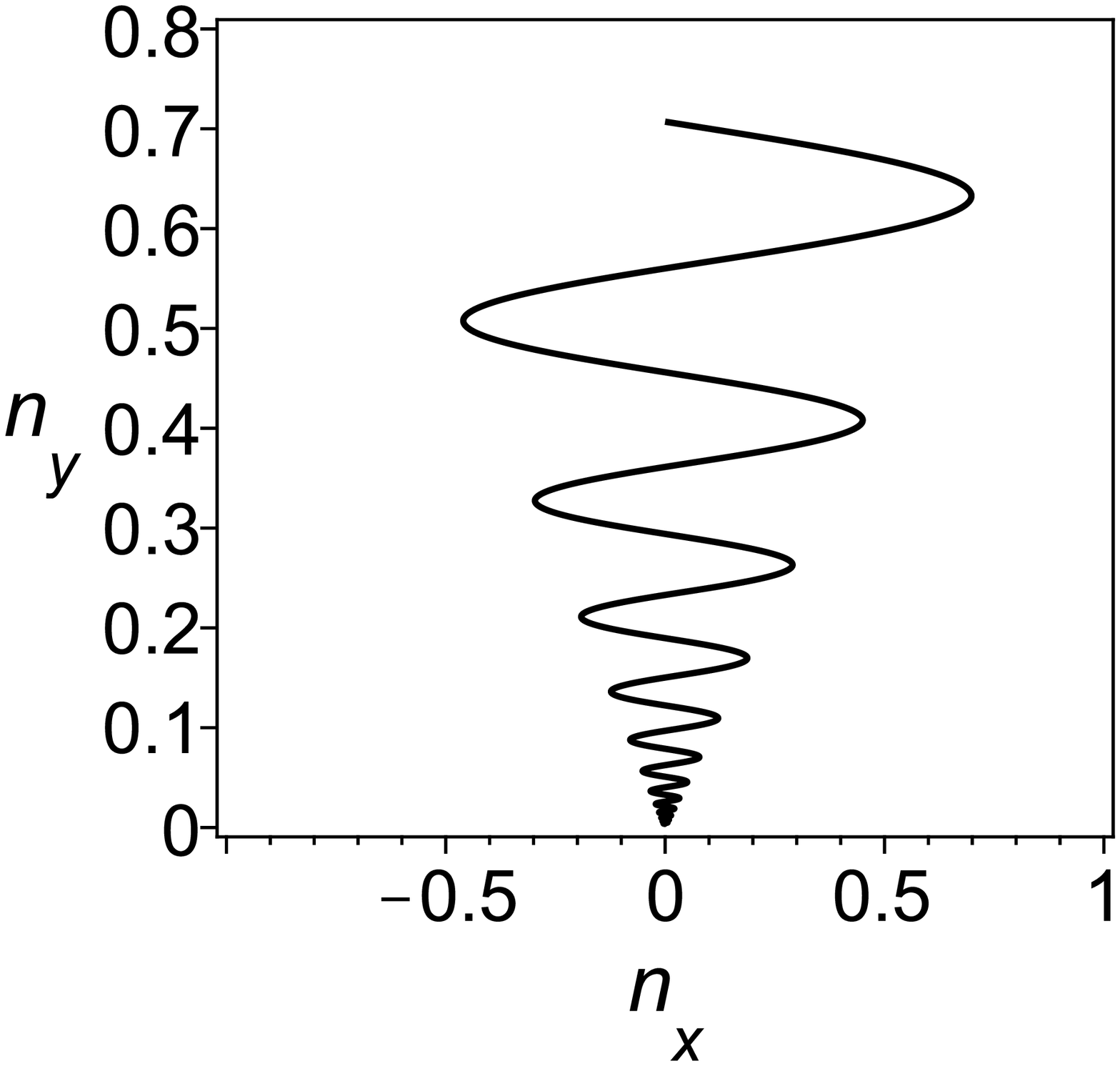}}
\scalebox{0.275}{\includegraphics{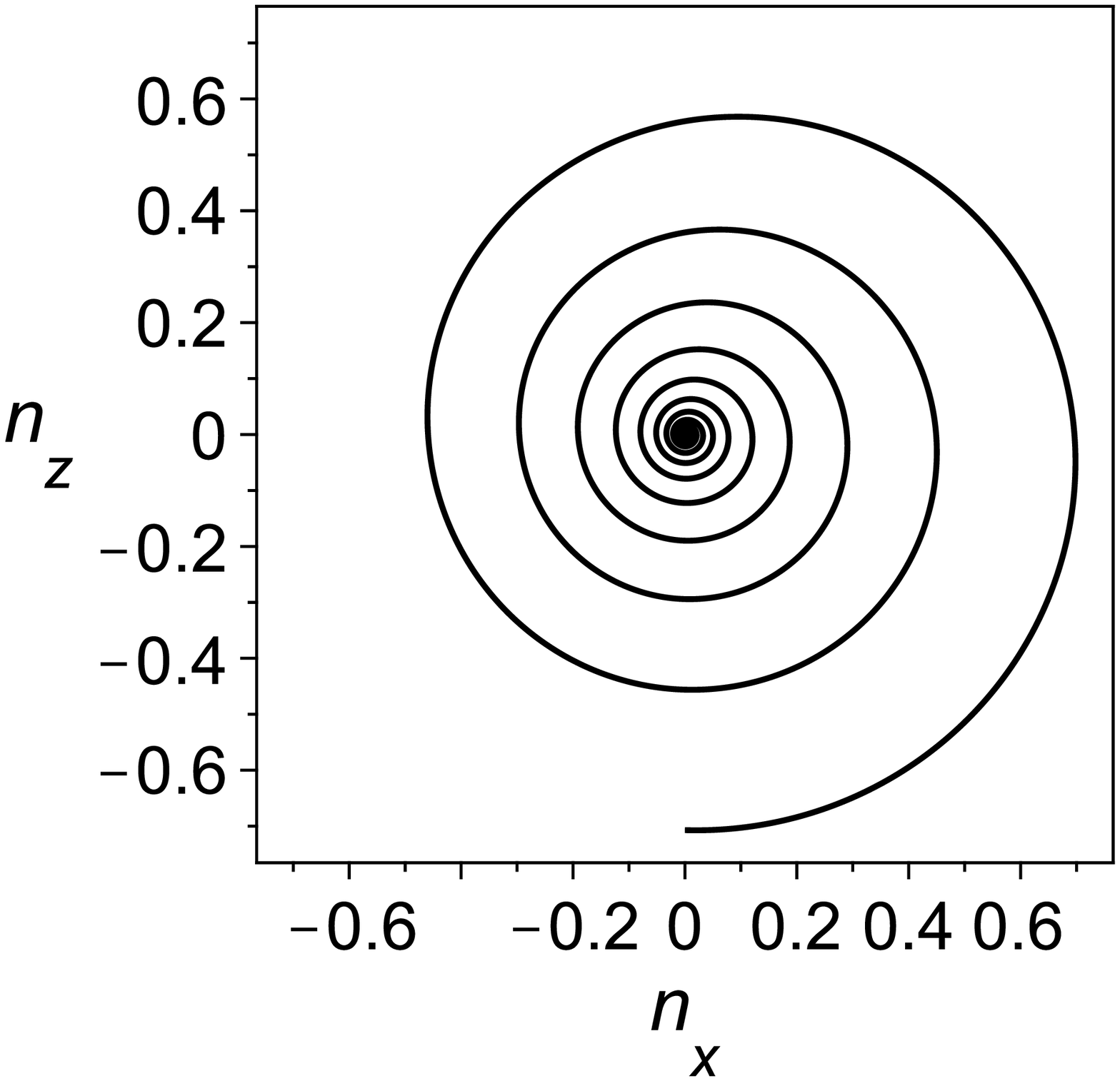}}
\end{center}
\caption{Projections of the Bloch vector on the \textit{xy} (left) and \textit{xz } (right) planes of the Bloch sphere. The parameters are $\Delta=0$, $\varphi=-\pi/2$ and the initial state is: $\mathbf n(0) =\frac{1}{\sqrt{2}}(0,1,-1)$. }
\label{BVxy}
\end{figure}

In Fig. \ref{BVxy} we present our analytical solution for evolution of the Bloch vector, $\mathbf n(t) = \langle u(t)|\boldsymbol\sigma |u(t)\rangle$.
The decay rate, $\Gamma$, and the on-resonance Rabi frequency, $\Omega_0$, are taken from \cite{SAM}: $\Gamma=0.035\, {\rm ns}^{-1}$, $\Omega_0/(2\pi) =80 \,{\rm MHz}$. Note, that the following relation between the density matrix and the Bloch vector, $\rho = ({n1\hspace{-.125cm}1}+\mathbf n\cdot \boldsymbol \sigma)/2$, takes place. Our theoretical predictions shown in Fig. \ref{BVxy} are in a good agreement with the experimental data and with the numerical calculations reported in\cite{SAM}.

\begin{figure}[tbh]
\begin{center}
\scalebox{0.4}{\includegraphics{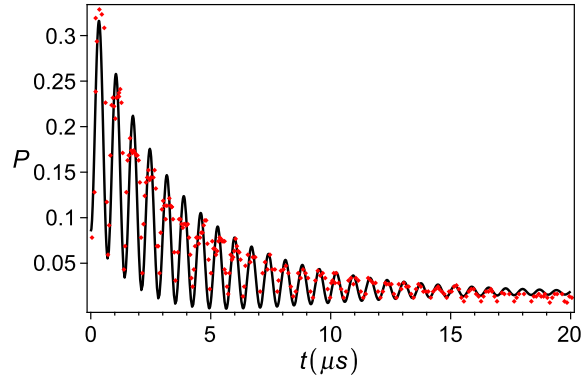}}
\end{center}
\caption{The probability, $P(t)$, of tunneling from the upper level to the continuum as a function of time. ($\Omega_0/(2\pi)=0.47 \,{\rm MHz}$, $\Delta/(2\pi)=1.34 \,{\rm MHz}$, $\Gamma = 0.204\, {\mu s}^{-1}$,  $\Gamma_{0}= 0.4\cdot 10^{-3}\, {\mu s}^{-1}$. The diamonds (red) are the data and the black solid line is the best fit to the oscillatory part of Eq. (\ref{P3a}). }
\label{PE1a}
\end{figure}

In Fig. \ref{PE1a} we compare our theoretical results with the reported experimental data \cite{YHC} on evidence for the Rabi oscillations in a current-biased qubit based on the Josephson junction. The Rabi frequency, $|\Omega|$, and decay parameter, $\Gamma$, are chosen as in \cite{YHC}: $|\Omega| = 8.9\cdot 10^6\,{\rm rad/s}$ and $\Gamma = 0.204\, {\mu s}^{-1}$. The initial state $|u(0)\rangle = 0.956 |0\rangle + 0.291 |1\rangle$, the detuning $\Delta$, and the on-resonance Rabi frequency, $\Omega_0$, are obtained by the fitting data to Eq. (\ref{P3a}). We see good agreement between the experimental data and the predicted theoretical behavior. We find that our time-dependent upper-level population is a better fit to the experimental data than the theoretical results presented in Ref. \cite{YHC} by Eq.(5). However, additional information on the experimental data is required for better comparison.

In the experiment reported in \cite{YHC}, the total tunneling probability was identified as the upper level population given by,
\begin{equation} \label{Eq2}
\rho_{11}(t)=e^{-\Gamma t} \frac{\Omega _{0}^{2} }{|\Omega |^{2} } \bigg|\sin \frac{\Omega t}{2} \bigg|^{2}.
\end{equation}
This is a particular case of our expression (\ref{P3a}), corresponding to the choice $C_1(0) =0$.

\subsection{ Tunneling into the continuum at zero ac current}

The fast frequency detector described in \cite{KA} is based on a superconducting phase qubit tunneling into continuum. In particular, the measurement is performed with ac current, $I_{\mu w}$, equal to zero.
Suppose, that initially the superconducting phase qubit is prepared as a superposition of ground and excited states.  Writing $|u(t)\rangle =C_1(t)|1\rangle +C_0(t)|0\rangle$, we obtain the time-dependent amplitudes $C_0(t)$ and $C_1(t)$ as solution of the Schr\"odinger equation
\begin{align}\label{eq12a}
&C_{1}(t)  =e^{-i\tilde\lambda_0 t/2} \bigg(\Big(\cos\frac{\Omega t}{2} - \frac{\Gamma - \Gamma_0+ i\omega_{10}}{\Omega}\sin\frac{\Omega t}{2}\Big) C_1(0)-   \frac{\Gamma_{01}}{\Omega}\sin\frac{\Omega t}{2} \,C_0(0)\bigg),\\
&C_{0}(t)  = e^{-i\tilde \lambda_0 t/2} \bigg (\Big(\cos\frac{\Omega t}{2} +\frac{\Gamma - \Gamma_0+ i\omega_{10}}{\Omega}\sin\frac{\Omega t}{2}\Big) C_0(0)-  \frac{\Gamma_{01}}{\Omega}\sin\frac{\Omega t}{2}\, C_1(0)\bigg ),
\label{eq12b}
\end{align}
where $\Omega = \sqrt{\omega^2_{10} -2 i\omega_{10}(\Gamma -\Gamma_0)- \Gamma^2}$ is the complex Rabi frequency. In the limit $\omega_{10} \gg \Gamma_1$, we have
\begin{align}\label{eq3}
|u(t)\rangle = e^{-i\omega_0 t }e^{- \Gamma_0 t/2}C_0(0)|0\rangle + e^{-i\omega_1 t} e^{- \Gamma_1 t/2}C_1(0)|1\rangle.
\end{align}
This simple formula is consistent with the known experimental data and the theoretical description of a partial-collapse measurement \cite{KNAB,KAB,PKA}.

The probability of escape from either  $|1\rangle$ or $|0\rangle$ is given by
\begin{align}\label{eq13a}
    P_{esc}(t) = 1- \rho_{11}(t) - \rho_{00}(t),
\end{align}
and in the same limit as above, $\omega_{10} \gg \Gamma_1$, we obtain
\begin{align}\label{eq13}
P_{esc}(t) = 1 -\rho_{11}(0) e^{-\Gamma_1 t} -(1- \rho_{11}(0))e^{-\Gamma_0t}.
\end{align}
As one can see, the tunneling into the continuum is described by a double-exponential decay function $P_{esc}$. This behavior of the probability of escape was observed in the experiment on measurement of dissipation-induced decoherence in a Josephson junction \cite{HYCW}.

While the contribution of the off-diagonal term, $\Gamma_{01}$, in the experiments discussed above is negligible, it can be more significant in the fast readout measurement of the phase qubit with a duration of measurement of approximately $\sim 3 ns$ \cite{CSM}.
\begin{figure}[tbh]
\begin{center}
\scalebox{0.3}{\includegraphics{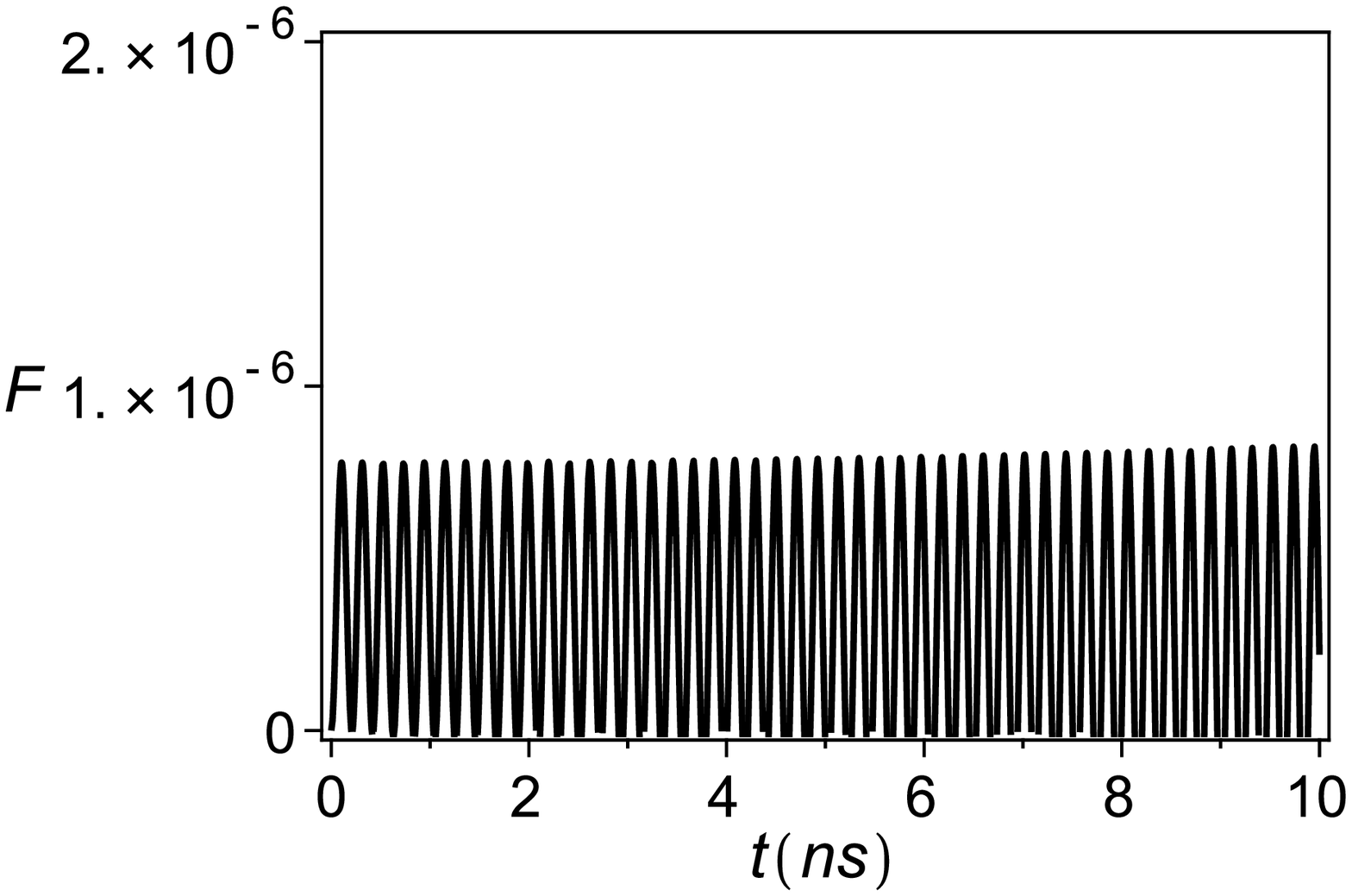}}
\scalebox{0.3}{\includegraphics{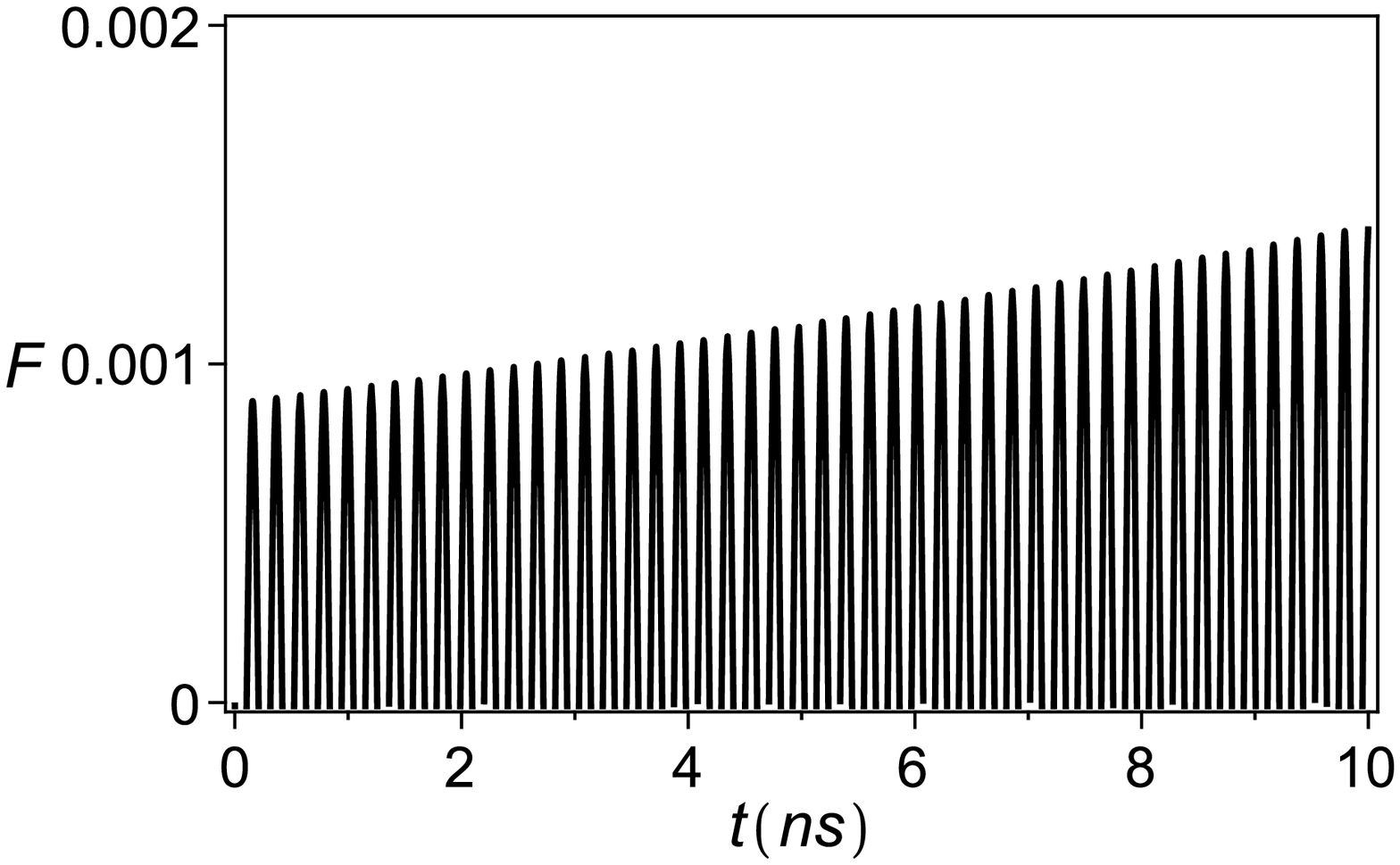}}
\end{center}
\caption{Deviation of population, $F(t)$, as function of time. Left: $C_0(0)=0, C_1(0)=1$. Right: $C_0(0)=C_1(0)=1/\sqrt{2}$.}
\label{F2}
\end{figure}
To estimate the contribution of the non-diagonal term, $\Gamma_{01}$, we introduce the function, $F(t)$, which characterizes the difference in probabilities when $\Gamma_{01} \neq 0$ and $\Gamma_{01} = 0$:
\begin{align}\label{F1}
    F(t)=\frac{\rho_{11}(\Gamma_{01},t)- \rho_{11}(0,t)}{\rho_{11}(0,t)}.
\end{align}
In Fig. \ref{F2} we present the results of computation for fast readout with the parameters being chosen as in  \cite{CSM}: $\Gamma_1 = 0.1\rm GHz$, $\omega_{10}/2\pi = 5 \rm GHz$ and $\Gamma_1/\Gamma_0 = 150$. One can see that, depending on the initial conditions, the contribution from the interaction between the two tunneling channels can be up to $\approx 0.1 \%$.

\section{Concluding remarks}

In conclusion, we have presented a non-Hermitian description of the dynamics of measurement of a supeconducting phase qubit. We have shown that our theoretical predictions are in a good agreement with the experimental data obtained in various experiments on superconducting phase qubits. We believe that the approach based on non-Hermitian Hamiltonians can simplify the theoretical description of the dynamics of the phase qubit measurement by tunneling to the continuum (in many cases and especially for a multi-qubit register).

This work was carried out under the auspices of the National Nuclear
Security Administration of the U.S. Department of  Energy at Los
Alamos National Laboratory under Contract No. DE-AC52-06NA25396. The work by AIN was supported by research grant SEP-PROMEP 103.5/04/1911. The work by GPB was partly supported by the IARPA.

\end{document}